# CloRoFor: Cloud Robust Forensics

Roberto Battistoni Dipartimento di Informatica
Sapienza Università di Roma
Rome, Italy
Email: battistoni@di.uniroma1.it
Roberto Di Pietro Dipartimento di Matematica
Universita Roma Tre
Rome, Italy
Email: dipietro@mat.uniroma3.it
Flavio Lombardi Dipartimento di Matematica
Universita Roma Tre
Rome, Italy
Email: lombardi@mat.uniroma3.it

*Abstract*—The malicious alteration of machine time is a big challenge in computer forensics. Detecting such changes and reconstructing the actual timeline of events is of paramount importance. However, this can be difficult since the attacker has many opportunities and means to hide such changes. In particular, cloud computing, host and guest machine time can be manipulated in various ways by an attacker. Guest virtual machines are especially vulnerable to attacks coming from their (more privileged) host. As such, it is important to guarantee the timeline integrity of both hosts and guests in a cloud, or at least to ensure that the alteration of such timeline does not go undetected. In this paper we survey the issues related to host and guest machine time integrity in the cloud. Further, we describe a novel architecture for host and guest time alteration detection and correction/resilience with respect to compromised hosts and guests. The proposed framework [1] has been implemented on an especially built simulator. Collected results are evaluated and discussed. Performance figures show the feasibility of our proposal.

## I. INTRODUCTION

Clouds offer a novel computing and storage model that allows performing large calculations possibly in parallel, enabling on demand service scaling following the number of requests, and possibly storing large amounts of data. All of this in an automated/autonomic setup without having to manage or pay the costs of a permanently deployed server farm.

Cloud computing is based on virtualization technologies. Multiple Virtual Machine (VM) guests can be dynamically deployed on a large number of physical hosts. This allows cloud providers to respond to resource request by on-the-fly (de)provisioning the required resources. Virtualization enables additional monitoring/inspection techniques but it is also a threat to guest data privacy [1].

Clouds also represent a big challenge for security and privacy. On the one hand they open the way to new cyberthreats and information security issues [1], which can be exploited by cybercriminals. In fact, an attacker can rent cheap cloud resources and apply techniques (e.g. brute force) that are now more affordable than ever. On the other hand the cloud model potentially enables more advanced computer forensics techniques, since cloud technology allows seamless retrieval and storage of Virtual Machine status (i.e. VM images) [2].

The cloud of clouds will be the next revolution in the cloud computing paradigm. Applications will then span multiple clouds over the network. As the complexity and size of clouds grows, it will be increasingly important to be able to protect such federated resources, especially from timing attacks. Further, with an anticipated growth of mobile users of cloud services in the near future, issues related to the timing of distributed and replicated cloud services becomes challenging. In particular, pricing models, trust and security-related issues have to be addressed.

A typical cloud or federated cloud scenario where multiple physical hosts run a large number of guest Virtual Machines, each with internal and external time sources is depicted in Figure 1. Cloud VM and service behavior are based on the timings they get through the internal virtual clock source exposed by the host. As such, time readings can be altered/-faked both by an attacker on the VM and by an attacker on the host. Time alteration attacks are quite common and well known in distributed systems [3]. However, cloud technology adds specificity to the problem [4]. Hence, the cloud scenario has to be modeled and investigated in order to improve the understanding of the problem and then to propose a remedy. In particular, prevention and detection of timing attacks are of uttermost importance in computer forensics [5].

### A. the Problem

In order to allow forensics in the cloud, getting the physical image of hard disks does not suffice [6]. As such, reconstructing the timeline of even a single user process could be very tricky even for a computer forensic professional (e.g. when guest migration is possible). This is true even if remote logging is in place, (e.g. having a remote rsyslog server) since cloud hosts and their guests can manipulate the internal timing used to send log data. The main threat we want to protect the cloud from is the malicious machine time variation in both hosts and guests. Such malicious alteration would undermine the efficacy of collected digital evidence for computer forensics investigations. In fact, if the timeline can be faked then the forensic analysis could be (at least) questionable.

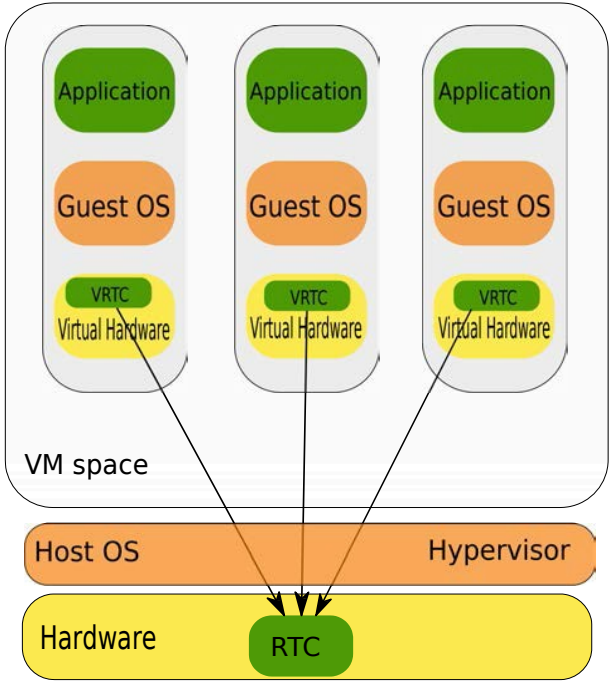

Fig. 1. Cloud computing scenario

The overall objective of this work is to improve resiliency of cloud hosts and guests in order to allow and ease forensics for the cloud, i.e. forensic activity aimed at investigating activity performed on the cloud by cloud and service providers and, by service users. Another goal is to gather as much information and insight as possible from simulating the possible attack scenarios.

### B. Contribution

In this paper we introduce a generic cloud time model and survey possible attack scenarios and countermeasures. We further provide a solution to the problem of protecting the timeline integrity (also referrelo to as "good timeline" in the following) of events inside guests and hosts. The proposed CloRoFor (Cloud Robust Forensics) architecture detects timeline alterations and ensures the validity of the forensics investigation process. We also show how cloud computing model and technology can help forensic investigation, how a malicious time alteration can be detected and how the original good timeline of events can be reconstructed. Both host and guest timeline alteration scenarios are described. As regards the latter, cloud snapshot and rollback techniques have been leveraged to bring the VM back to the original unaltered state.

The remainder of this document is organized as follows: Section II summarizes state of the art solutions applied to cloud resilience; Section III introduces a generic time model for cloud computing; Section IV shows the proposed CloRo-For architecture; Section V gives implementation details and depicts the features of the novel cloud simulator that was implemented; Section VI presents performance figures and discusses results; finally, in Section VII conclusions are drawn.

## II. RELATED WORK

Cloud security is of paramount importance today, but not much attention has been devoted to cloud timing attacks as to more traditional server timing attacks in the past [7], [8], [9]. Much recent efforts have been especially devoted to address cloud privacy issues. Work by Van Dijk [10] shows that fully homomorphic encryption (FHE), cannot enforce the privacy required by common cloud computing services. As such, cloud service security has to rely on additional privacy enforcement tools, such as tamper-proof hardware, distributed computing, and complex trust ecosystems.

Further, the time taken by web sites to respond to HTTP requests can leak private information, using different attack types. Direct timing, measures response times from a web site to expose private information [8] such as validity of a username or the number of photos in a public gallery. Cross-site timing, allows obtaining information from another site [7].

Digital forensics in the cloud is quite a hot topic that has been investigated by Birk [11] and Franc [12]. The former analyzes and addresses technical issues of cloud forensics whereas the latter attempts to use existing tools to collect data from the cloud and gives suggestions and tools on the subject. Also Dominic [13] discusses cloud forensics and stresses the decentralized nature of data processing in the cloud. Dominic focuses on the technical aspects of digital forensics in distributed cloud environments and assesses whether it is technically possible for the cloud computing service customer to perform a traditional digital investigation. Marty [14] introduces a logging framework for the cloud to ensure that the data needed for forensic investigations has been generated and collected. Marty's approach is complementar to our proposal in that focuses on easy and effective log collection. However, this approach does not protect against timing attacks such as the ones addressed here. Cloud forensics and privacy issues have been dealt with in [15] using homomorphic encryption [15] and commutative encryption. Two forensically sound schemes are provided so that the investigators can obtain the necessary evidence while the privacy of other users can be protected.

Several papers in the literature have addressed anti-forensic attacks, i.e. attacks aimed at preventing forensic analysis such as timing attacks. However, no general frameworks were developed to counter such attacks. A taxonomy of anti-forensic attacks and countermeasures is presented in [16]. An interesting work on live digital forensics addressing rootkits that is complementar to our work is by Carrier [17]. However, to the best of our knowledge, no previous work on cloud forensics has modeled, studied and addressed timing attacks in depth as present work.

Another hot topic is mobile cloud forensics. This has been initially investigated by [18] introducing a service concept called "Forensic Cloud" and suggesting a simple technology framework for forensic analysis based on a mobile cloud.

Finally, there is previous work regarding adequate laboratory specifications for emulating network attacks and experimenting with network forensics [19]. The findings of such work validate our testbed implementation.

## III. CLOUD TIME MODEL

This section introduces a first high-level time model that is used to study, evaluate and discuss potential time-based abuses, threats and possible remedies. In particular a taxonomy of attacks and possible countermeasures is given. We aim to guarantee the timeline integrity of guest services. In this first model we assume a central unique remote trusted and protected entity, namely a Cloud Controller (CC in the following, see Section IV), that cannot be altered or faked by an attacker. **Agents** are software units distributed in every cloud machine which manage their time and cloud time.

**DEF: Good Timeline**: We define a good timeline to be a series of events ($e_i$) in a machine that have strictly increasing associated timestamps ($ts(e_i)$) for events happened one after another.

**DEF: Guaranteed Timeline**: We define a Guaranteed Timeline as a series of events with associated timestamps, where we guarantee that for every event $e_x$ $e_y$, $ts(e_x) < ts(e_y)$ iff $e_x$ happened before $e_y$ [20].

### A. Abuse Cases

Some of the possible "Abuse Case" scenarios that alter the normal "good timeline" are as follows:

**Abuse Case 1**: an attacker obtains admin privileges on a *guest* VM. She changes machine time in order to alter and fake the timeline (of events).

**Abuse Case 2**: a malicious user has admin privileges on a cloud *host* and can change host and consequently guest machine time at will.

In order to guarantee a good timeline to N distributed systems hosted on M possibly malicious hosts, time information exchange, action and reaction systems have to be put in place. Different approaches can be taken:

DTS Distributed Time Synchronization: time information can be distributed in P2P fashion and decision on the right timing is taken with a quorum approach; This approach can be distributed across different clouds but suffers from traffic explosion issues [21];

HTS Hierarchical Time Synchronization: periodic heartbeat timestamps (see Section IV-D) are broadcast from one or more secure CC to all hosts and their guests. Hosts must appropriately route such timing packets to all contained guests. Scalability issues exist if a reverse channel is used where all guest and host agents are required to send periodic heartbeat timestamps back to the secure Time Stamp Service (TSS, see Section IV). This can cause feedback implosion issues, that can be handled in different ways [22].

A set of essential properties have to be guaranteed by a good timeline guarantee/verification system:

- Accountability: based on the received timestamps, Time Controllers (TC, see Section IV) have to record timestamp evolution over time.
- Resiliency: a guest has to be resilient against one or more malicious hosts and vice-versa. Of course it is more difficult for a guest to be resilient to host-based attacks. Details can be found in tables I,II,III,IV.

### B. Scenarios

*1) Secure Channels with Unforgeable Agents:* A secure channel is defined as a channel protected by a symmetric session key eventually augmented with an asymmetric handshake phase. A number of key management solutions can be deployed. Encryption keys are protected with anti-forging mechanisms leveraged by agents (see code obfuscation and keys below [23]).

*2) Secure Channels with Forgeable Agents:* The ultimate attack detection decision is due to the guest agent, such decision has to be implemented using a conditional jump instruction that the host can intercept. The goal here is to prevent the host from intercepting conditional jumps or test instructions inside the guest VM [24] [25]. This is difficult since the host platform can access all physical and virtual resources of guest VMs. Code obfuscation [26] can help protecting the guest from host interception. In order to achieve that, the code can obfuscatedly compute a function [27] of a large input data (e.g. different time probes taken in different ways from the agent). As such, the output will not be a simple true/false result but a combined result that only the destination CC can validate and extract information from (optionally the sent data can be signed by the sender) [28].

### C. Assumptions

One agent exists for each guest and for each host. Every agent is implemented using a different code obfuscation method and hides its own symmetric key or an asymmetric key in the code. All guest and host agents symmetric keys or counterpart asymmetric keys are known to the CC (that can be a single point of failure).

### D. Time Resiliency with Unforgeable Agents

**Guest Time Resiliency**
This is the case of "Guest Time Resiliency assuming Agents are not forgeable" (see Table I).

**Host Time Resiliency**
This is the case of "Host Time Resiliency assuming Agents are not forgeable" (see Table II).

### E. Time Resiliency with Forgeable Agents

**Guest Time Resiliency**
This is the case of "Guest Time Resiliency assuming Agents are forgeable" (see Table III). In this case it is very difficult to detect whether a non-malicious guest is reporting some violation or the guest agent is actually malicious and as such it cannot be trusted. Further, the host can forge any message or response coming from any hosted guest, given that in this scenario, keys in agents cannot be ultimately protected. This leads to alternative solutions where probing the actual time has to be done indirectly:

- creating interactive service requests and analyzing the obtained responses to infer internal service times;
- capturing and analyzing actual service request and response packets in order to passively infer internal service times.

| Table | Malicious Guest | Non Malicious Guest |
|---|---|---|
| **Malicious Host** | Both Host and guest agent detect alteration and inform CC. Heartbeat reveals possible shutdown of Agents. Guest migration could improve the resiliency of the Guest machine | Agent on the guest can notify the CC that Host is compromised. CC takes countermeasures |
| **Non Malicious Host** | Host immediately intercepts [25],[1] guest attempts to modify clock and prevents them: guest agent exposes a service to get guest time and then compare that with the Host time | Ok by definition |

TABLE I
UNFORGEABLE AGENTS: GUEST TIME RESILIENCY

| Table | Malicious Guest | Non Malicious Guest |
|---|---|---|
| **Malicious Host** | Both Host and guest agent detect alteration and inform CC. Heartbeat reveals possible shutdown of Agents. | Guest Agent can probe host services and detect host time has been altered: can notify the CC that Host is compromised. CC takes countermeasures |
| **Non Malicious Host** | Malicious guest cannot affect Host Time | Ok by definition |

TABLE II
UNFORGEABLE AGENTS: HOST TIME RESILIENCY

Such probing has to be done externally to the malicious entity. If services are not interactive (i.e. request/response) but long-running batch jobs [29], then there must be a way of interrupting them in order to get partial responses. Of course smart semantics-aware packet sniffing is a requirement for this kind of analysis.

Agent protection is based on code obfuscation [30] [31] [32]. As in [33] [34] the goal is to be able to perform a similar computation that yields probabilistic values. Such probabilistic results can always be decoded by the CC in order to infer a true/false result. It is however possible to compute an intermediate result based on embedded key input (and a shared time). Based on such encoded partial result it is possible for a CC (who has counterpart key) to extract a true/false final result. Such final function is extremely difficult to compute without a pre-shared secret (and a shared time) [28]. Hence we obtain a function that is difficult to reverse-engineer and to emulate.

**Host Time Resiliency**

This is the case of "Host Time Resiliency assuming Agents are forgeable" (see Table IV).

## IV. CLOROFOR ARCHITECTURE

This section describes the CloRoFor architecture for protecting cloud guests and hosts from possibly malicious time alterations. Definitions of the most important CloRoFor components and of their scope and interactions is given. The main architectural goal of CloRoFor is to provide a framework where selected components can be deployed in order to detect and react to attacks and anomalies in the cloud timelines.

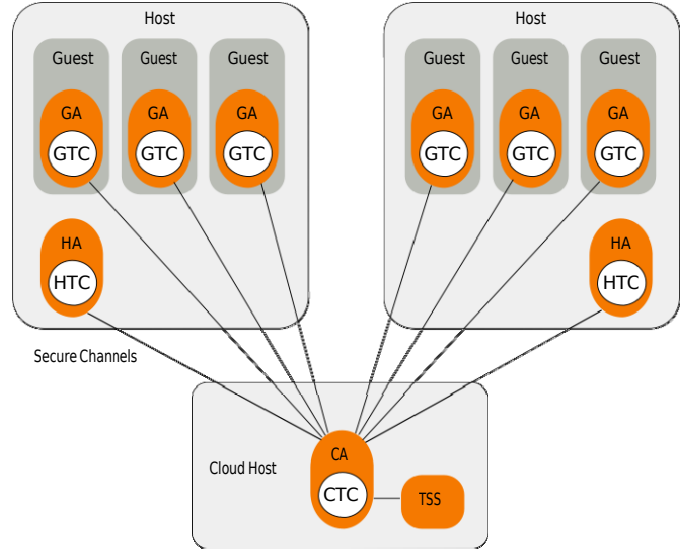

Fig. 2. Proposed Architecture

In Figure 2 the proposed architecture is depicted. The main CloRoFor component is the local Agent. The Agent is an abstraction comprising all actors (TSS, CTC, GTC, HTC) showed in the following.

### A. Time Actors

Time Actors are the abstract model of the entities which contribute to the Cloud Time Model. Every actor is implemented as an Agent and deployed into a physical/virtual machine of the Cloud.

### B. Software Agents

In this section we give further details on software agents, relating them with security aspects depicted in IV-D:

| Table | Malicious Guest | Non Malicious Guest |
|---|---|---|
| Malicious Host | A malicious guest can pretend behaving well when on a non malicious host. NO detection | Malicious Host can forge both host and guest agent. so NO detection |
| Non Malicious Host | Host detects guest time alteration: host could intercept guest network packets [1], understand packets type and compare packet timestamp with Host time. But guest agent might lie and tell CC that host is malicious. Then CC has to choose. | Ok by definition |

TABLE III
FORGEABLE AGENTS: GUEST TIME RESILIENCY

| Table | Malicious Guest | Non Malicious Guest |
|---|---|---|
| Malicious Host | totally NO Detection | Host can forge guest agent. NO detection |
| Non Malicious Host | Host detects guest time alteration but guest agent might lie and tell CC that host is malicious. Then CC has to choose | Ok by definition |

TABLE IV
FORGEABLE AGENTS: HOST TIME RESILIENCY

- Def (Guest Agent, GA): every Guest has a software Agent whose code is obfuscated and contains symmetric and/or asymmetric keys.
- Def (Host Agent, HA): every Host has a software Agent whose code is obfuscated and contains symmetric and/or asymmetric keys.
- Def (Cloud Controller Agent, CA): the Cloud Controller has a software Agent whose code can (optionally) be obfuscated and contains symmetric and/or asymmetric keys.

The core of the agent is the *TimeController (TC)* entity that monitors the machine timeline and delivers and receives heartbeats.

### C. Time Controller

- Def (Time Stamp Service, TSS): is a service that updates local time from one or more reliable external timing sources and it is built into the CC;
- Def (Cloud Time Controller, CTC): invokes the TSS in order to update the Cloud controller (CC) time; the CC features a local agent called CA which implements the CTC.
- Def (Guest Time Controller, GTC): traces the Guest timeline and sends such traces to the CC; for every Guest a local agent exists called GA implementing a GTC.
- Def (Host Time Controller, HTC): the Host Time Controller is analogous to the GTC but for the host time and the HA is the HTC implementing agent.

### D. Key Security Considerations

We assume that host and guest agents communicate with the CC agent through a secure channel. Said that, we can discuss the most important security aspects:

**Time Heartbeat**: Every guest and host sends a periodic *heartbeat* (HB) packet containing timestamp information to the CC through the above-defined secure channels. This one is needed to get the guest/host status and timestamp at every moment (following the sync properties seen before). The nature of the HB is dual: it communicates the existence of the peer and its timestamp.

**Anti-forging mechanisms**: Assuming that agents are unforgeable, the mechanisms to implement that, it is via code obfuscation. An example of this technique is Skype client that makes use of various tricks like junk code to protect encryption keys and code from the reverse engineering technique ([35], [36]). Agent code (and keys inside it) must be frequently updated to minimize the probability of reverse engineering.

**Key Management**: It is needed only in the case Agents are unforgeable because, in the case the Agents are forgeable, we must consider the keys lost and in the hands of the enemy. The keys are embedded in the Agents and protected by obfuscation mechanisms (see [27]). Said that we have two cases:

- **Symmetric keys**: the CC has to store all host and guest keys (this would bring scalability problems [37]);
- **Asymmetric keys**: every guest has to store CC public key and the CC has to store public keys for all guests and hosts; Guests, hosts and CC exchange symmetric session keys through an asymmetric handshaking phase. Session keys are used to encrypt the above defined channels;

### E. Cloud Oracle

The cloud controller (CC) is the single unique trust object in the cloud and it finally decides about ambiguous use cases such those in the Section III-C. For example it answers the following questions:

- Both host and guest are signaling each other as malicious: which one is really malicious?
- Many hosts have been signaled as malicious: are they really malicious? Or are the ones that signal them?

Out of the first Proof of Concept is the Cloud's Oracle and other details that are not relevant to the overall result.

## V. IMPLEMENTATION

This section provides details over the implementation of a first prototype of the proposed architecture. This includes the cloud simulator that has been especially designed to test and stress as much as possible clouds composed of real nodes.

In this work we implemented a Proof of Concept of the Agent. We chose to implement the prototype in Java for the following reasons:

- the high level language introduces some overhead which can help simulate the effect on performance of an obfuscation function;
- the language offers the possibility of extending the simulation on more than one machine;
- the language allows for fast implementation tuning and evolution.

The obfuscation of the communication channel is an important part of the security model. The reason why is that detection/security-related information has to be kept as hidden as possible from an attacker. In fact, we chose not to leverage well known protocol implementations such as SSL, since it would have made easier for an attacker on the host side to infer information from the guest VM. We also chose not to use the TCP protocol because it would not have been efficient for HB message exchange (as it would have required connection establishment at each HB exchange). TCP was discarded also because using UDP for all communications is potentially less prone to tracking [38]. So we adopted a datagram protocol such as UDP and our own implementation of the secure channel and compromising signaling.

**Additional simulation components**: in order to model the behavior of a particular action/functionality request, we implemented in a class SimulationModel a method askOracle() that implements the desired behavior (e.g. failure probability according to a chosen distribution). Every node and object instance can instantiate a different SimulationModel (SM) object, thus behaving in a different way. Groups of nodes/object instances can instantiate/refer to the same SM object and thus we can model complex system interactions failure probability.

**Time controller and TimeStamp Service**: The main implementation objects are the Time Controller (and its declination for Hosts, Guest and Cloud Controller, HTC, GTC and CTC) and the Time Stamp Service (TSS). A TSS on the cloud controller is different from the TSS on Guests or Hosts because it could have a connection to an external source to get a faithful and trusted time. In this implementation we consider the time of the Cloud Controller as trusted, so there are not differences through TSSs. HTC and GTC share the same implementation as they behave in the same way.

**Secure channel**: As a first PoC, we developed a symmetric ciphered channel for channel confidentiality. The asymmetric case is an extension of this one. The peer agents know the symmetric key and encrypt the Datagram packets with it. Given that UDP is not reliable, we could introduce redundancy in transmissions, sending burst of the same packets repeated for N times, in order to minimize the loss probability. However, using bursts does not solve the packet loss issue. On the contrary, it makes it worse. This is the reason we simply do not try to recover lost datagrams. On the contrary, we see the lack of reception of a timestamp packet as a security warning. If the lack of datagram from a source node continues we can increase consciousness that something bad happened, be it network congestion, malicious host intervention or guest inability to send (crashed or victim of a successful attack). For the authentication of the peers the mechanism used is the challenge/response with keys that are hard coded in the obfuscated code. As a consequence, keys are renewed by updating the code containing them.

**The Heartbeat channel**: Every message, such Heartbeat or Heartbeat+Signal message, has an ordinal attribute to build the timeline of the sent or received messages. That ordinal number allows the analyzer to check for anomalies, like those where for growing ordinals there are not corresponding growing timestamps.

We chose not to separate the signaling (anomalies, etc) channel from the regular HB channel. This is so in order to protect as much as possible the signaling from tampering and blockage. In fact a malicious host might intercept spurious datagrams not sent at regular intervals. In fact, an irregular sending time would possibly mean to the CC that the packet contains some other kind of information. A malicious host would most probably try to suppress such messages as the CC would not notice that packet is missing. Hiding the signaling channel inside regularly-spaced HB timestamp packets has advantages and disadvantages:

ADV1 : with covert channel the attacker has to decrypt and semantically interpret the datagram content in order to make use of it;

ADV2 : the signaled event/status can be tied to a particular point in time and as such replay attacks are not possible;

DIS1 : any signaling of anomalous conditions/alarms is deferred up until the following heartbeat packet departure, this can potentially allow the attacker to perform further damage.

Heartbeat packets are in the form *HB(challenge, signal payload, timestamp)* for the those packets sent from the CC to the Guests/Hosts and *HB(response, signal payload, perceived time)* for those sent from the Guest/Hosts to the CC. The Cloud Controller (CC) verifies timelines and can alert an external entity or can directly warn the administrator. However, such signaling from guest or hosts to the CC has to be protected as it can be exploited by the attacker to induce false detections. As a countermeasure against man in the middle attacks, the datagrams that are sent contain ordinal sequences as described above.

### A. A Cloud Simulator

We designed and implemented a software simulator (it will be soon released as open source) that could allow performing experimental activity over the CloRoFor solution we devised. The simulator models a cloud scenario with multiple hosts each running a number of guests. The simulator runs inside a JVM and actually implements the software components that are part of the CloRoFor architecture. The simulator executes agents as software threads that actually perform computation over data coming from actually sent and received datagram packets. This way small scale results can be obtained but also validated in a real testbed. The simulator also allows performing wide scale experiments before actual large scale deployment. We decided not to make use of existing cloud simulators for a variety of reasons:

- The cloud simulator Cloudsim [39] did not allow to actually implement and study the behavior of the required functionality and to collect packet loss statistics;
- The network simulators ns-2 and ns-3 [40] would not allow to simulate complex interactions among cloud guests and hosts.

## VI. EXPERIMENTAL RESULTS AND DISCUSSION

Present section describes the experimental activity performed on the prototype and aimed at extracting useful information on the behavior, scalability and limitations of the proposed approach. This section also aims at measuring the impact of a number of malicious/lost nodes on the overall cloud.

The experimental activity aims at evaluating the capacity of our system to detect node compromising and, as a consequence, react and restore security. The robustness to attacks can be measured using the following indicators in this scenarios:

- Malicious Attack: Erroneous Timeline, Erroneous Response, Heartbeat miss.
- Packet Loss/Delay: Host congestion, Network congestion.
- Node failure: HeartBeat Loss, Heartbeat Response. Over

the implemented CloRoFor prototype actual functionality and performance tests have been performed and results shown and commented in the present section.

### A. Test Plan

Tests have been performed over a virtual Cloud of seven virtual machines, each with a Linux-based OS (Fig. 3). Every VM has 512 MB of RAM, Linux, Java 1.6 JVM, 1 Processor assigned to the VM. Every VM is a Cloud Host: the first one only contains the Cloud Controller jthread (Java Threads), the other ones the sets of Guest jthreads. Every Guest runs an Agent with its Time Controller (TC) executing inside it.

Running the current CloRoFor version implies having a maximum of 400 simulated guests (as jthreads). Since HTC and GTC are the same, we count HTC as TC. So the 400 TC are composed of GTCs and of one HTC. The value 400 is an upperbound for TCs because it corresponds to circa 2400 VM's OS threads (every Agent implementation features around 6 VM threads for each one Agent jthread), which is the maximum limit supported by the standard Linux kernel. However, we consider this value a reasonable upperbound that does not affect simulation results.

Every CloRoFor instance has a configuration file in which CloRoFor parameters are defined. For this first series of tests we configured them as in Table V. Some considerations are needed about config parameters. As regards guest failure percentage, in general, guest VM failures are induced by the failures on their host. Other guest failures can only be due to virtualization issues that can affect guest behavior. Said that, we assume the cloud system as highly reliable, and the assumed failure percentage exclusively due to guest failure has been set to 0.001%.

As regards delay timing, we configured the system in order for an HB packet to be sent every 30 sec. from a Guest to the CC (through the Host). The CC checks the received packets every 30 sec. for response and timeline correctness, and every 5 min. for the timeout threshold. It is worth noting that by "Timeout" we mean that the packet arrives to the CC but only after the expiration of the 5 min. timeout threshold (ex: a packets arrives after 5 min. and 10 sec.).

In the tests we varied the overall running time of CloRoFor simulation. Every single test was repeated at least 4 times. Running time was set to: 30, 60, 120, 240 and 480 minutes (following a quadratic growth law to minimize the trails number and maximize test significance).

HBs arriving to the CC are analyzed in order to find *Timeline-*, *Timeout-* or *Response-errors*. Errors are possibly introduced by infrastructure problems (failure probability) or by attacks (hacking probability). An HB with errors is captured and then deleted from the CC data structures to limit the simulator footprint in memory. As such, it is not presently possible to trap HBs with more than one error. This limitation of the simulation model is induced by the implementation of this first prototype. Trapping more than one error for HB would have meant maintaining large data structures for too long thus limiting simulation running time. In this first Java prototype we did not focus on memory optimization but only on the interesting insights and results we could get from this first series of simulations. An extension to this work could be to maintain the trapping history (instead of packet history) of every guest/hosts to minimize that memory footprint. However, we consider the present prototype and test setup to be more than adequate for the kind of simulations we were interested in performing.

### B. Simulation Results

In Figure 4 and 5 we plotted the number of Timeline-, Timeout- and Response-error packets and the false positive (FP) percentage of these kind of errors (notice the x-axis scale). The FP shows how many error packets are due to the infrastructure (simulated) failure instead of the hacking (simulated) activity.

Response errors FP (Fig. 5) are always zero and this is natural because there is no infrastructure problem (we assume the transport layer does not produce any malformed UDP packet) that can induce a response error. Only malicious hacking activity on the guest or host can induce them. As such, we simulated them in these two test sets.

Timeout errors in Fig. 4 (with the meaning we explained above) are low if the running time of the simulator is short and increase (exponentially) as the simulation time enlarges. This seems natural since the longer the time of computation, the higher is the number of packets that are traveling. The exponential growth of the packets and the FP%s tends to 100%. The reason why is that system performance degrades up to a point where it cannot cope with packet analysis anymore.

Timeline errors FP (Fig. 5) can only be observed when the system is heavily loaded. This is a very interesting phenomenon because the system is under pressure and this increases the time required for an HB to arrive to the CC. Further, it increases the probability that some packet will arrive out of order at the destination. This is surely related to our decision to implement the communication layer using the UDP protocol instead of TCP.

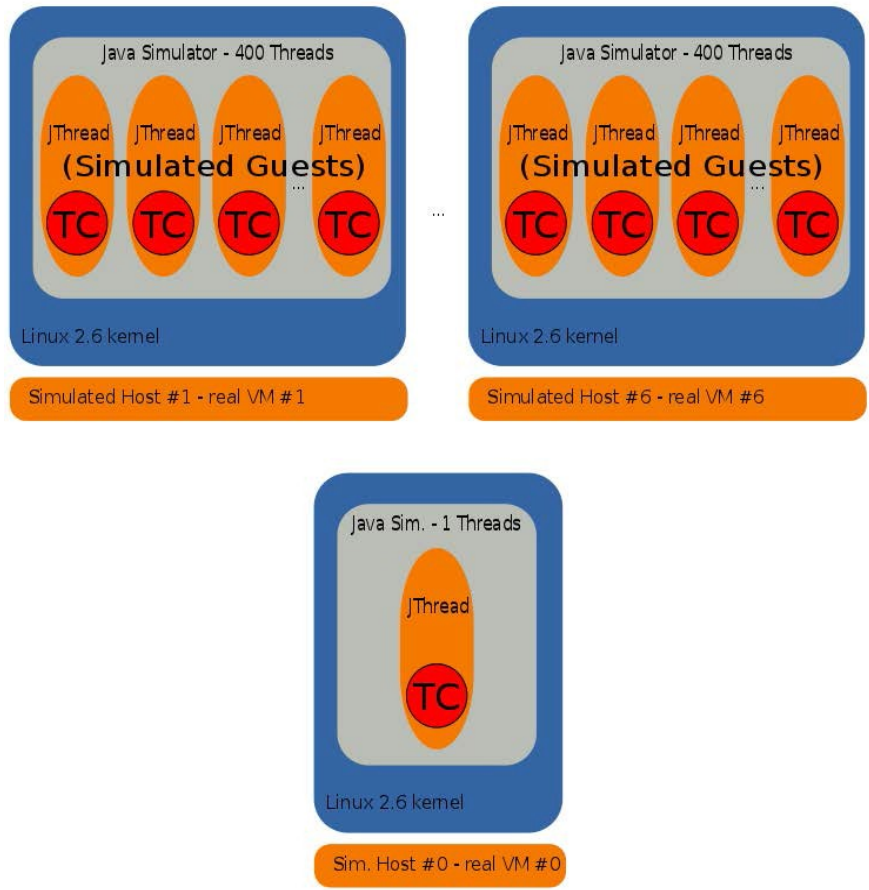

Fig. 3. Testbed: simulated Hosts and Guests

In order to better understand/show the evolution over time of the system, in Figure 6 we plotted the frequency distribution over time of Timeout-, Timeline- and Response-errors. It appears on the one hand that Timeline- and Response-errors follow very similar linearly decreasing curves (as the x-axis is logarithmic in time). Same similarity is present in Fig. 4. This seems natural since the environmental conditions of the simulations were the same for these two phenomena. The above curves are linearly decreasing, this means that Timeout-errors phonomenon eventually dominates the other two phenomena. As such, the simulated system shows

Timeline- and Response- errors appear as affected by timeout-errors more than other. On the other hand, the Timeout error frequency line, slightly decreasing at first, shows an exponential growth for larger x values. This can be due to a saturation effect in the system due to the fact that additional delays add up in the system (same as in the Fig. 4).

The most evident result from the experimentation is that the FP Timeline is low even when the FP Timeout increases. This shows our implementation is accurate for Timeline because it preserves Timelines from infrastructure errors. At the same time it emerges that the our Java implementations poses too much overhead on the network and this is reflected by the Timeout-errors high number and high FP which limits the simulation duration.

| Parameter Name | Value | Notes/Explanation |
|---|---|---|
| **HeartBeat delay sending time** | 30 sec. | the value is reasonable to not overload the whole system while having a reasonable sending time too |
| **CC checking time interval for packet respons errors** | 30 sec. | checking time is aligned to the sending time |
| **CC checking time interval for packet timeline errors** | 30 sec. | checking time is aligned to the sending time |
| **CC checking time interval for packet timeout errors** | 5 min. | after minutes of time you can consider the arriving packet as an overtime packet. So Timeout is reached |
| **Guest failure probability** | 0.001% | Guests do not fail, but representative percentage has assigned |
| **Host failure probability** | 2% | Hosts fail for many reasons and considering their often better quality hardware the fail probability is very low |
| **Guest probability of being hacked** | 5% | It is not so common to be hacked So the hack probability it has to be higher of fail probability but not too much. |
| **Host probability of being hacked** | 2% | Hacking an Host requires very high skill and knowledge of the infrastructure because the Host is not exposed as the Guests in it. So we assigned the same probability of the fail case |

TABLE V
SIMULATION PARAMETERS

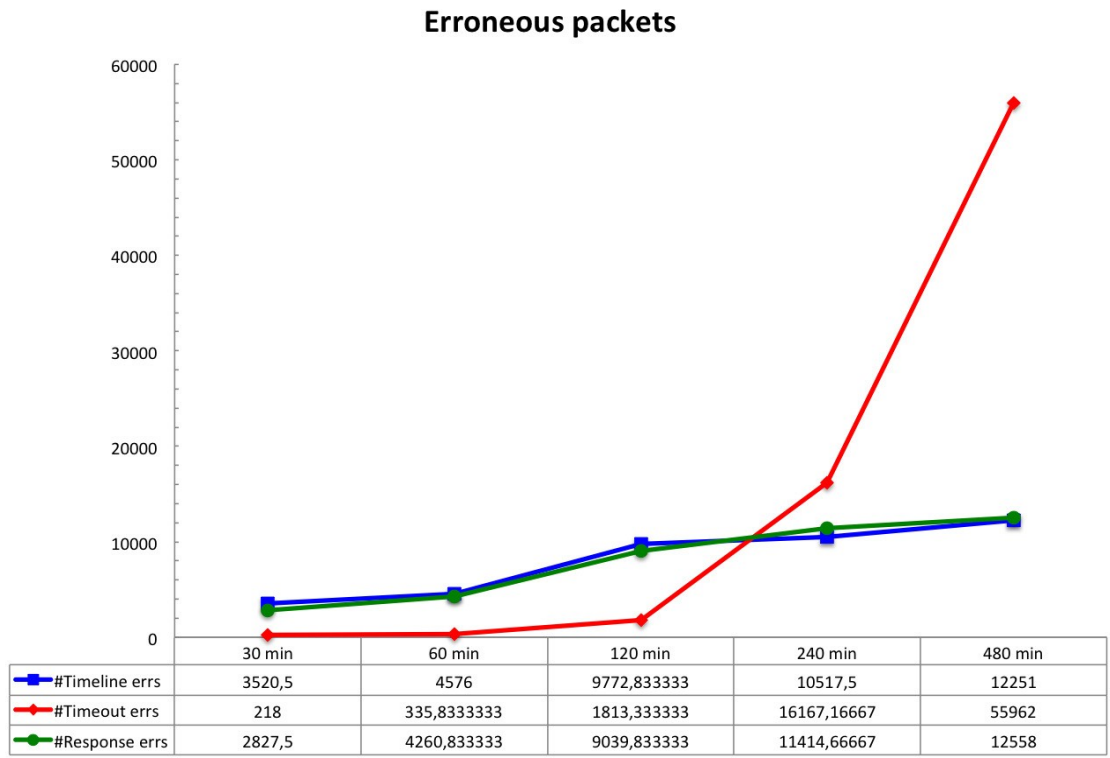

## Erroneous packets

| | 30 min | 60 min | 120 min | 240 min | 480 min |
|---|---|---|---|---|---|
| #Timeline errs | 3520,5 | 4576 | 9772,833333 | 10517,5 | 12251 |
| #Timeout errs | 218 | 335,8333333 | 1813,333333 | 16167,16667 | 55962 |
| #Response errs | 2827,5 | 4260,833333 | 9039,833333 | 11414,66667 | 12558 |

Fig. 4. Simulation results: erroneous packets for Timeout, Timeline and Response

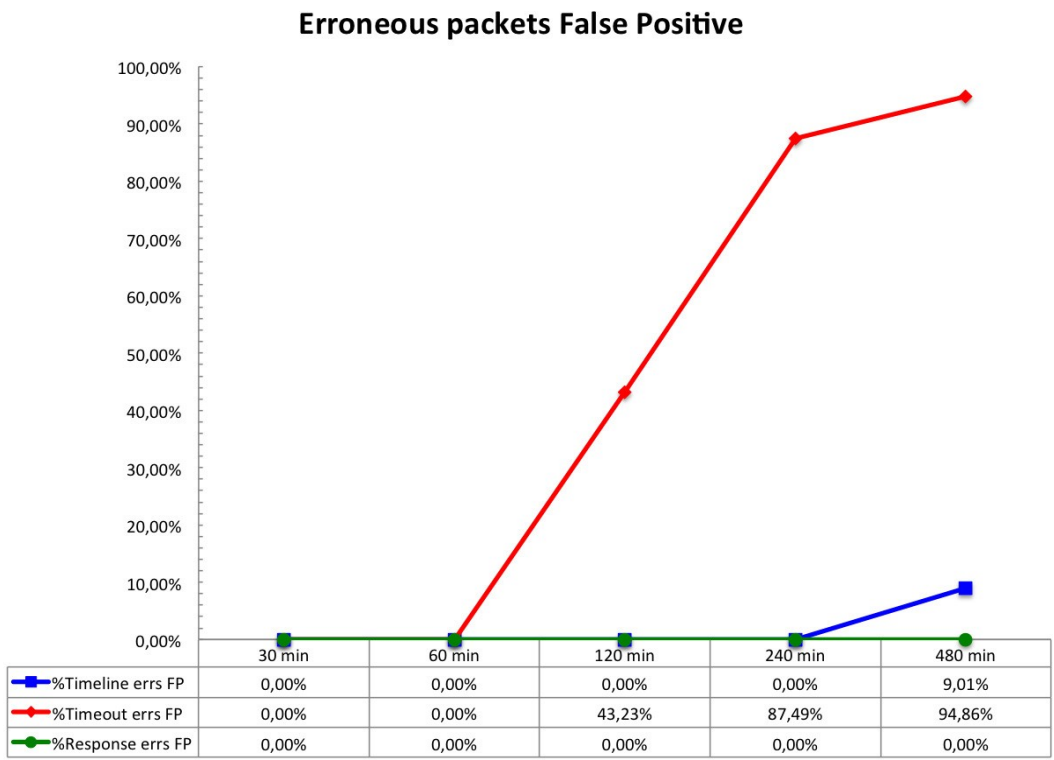

Fig. 5. Simulation results: erroneous packets false positive for Timeout, Timeline and Response

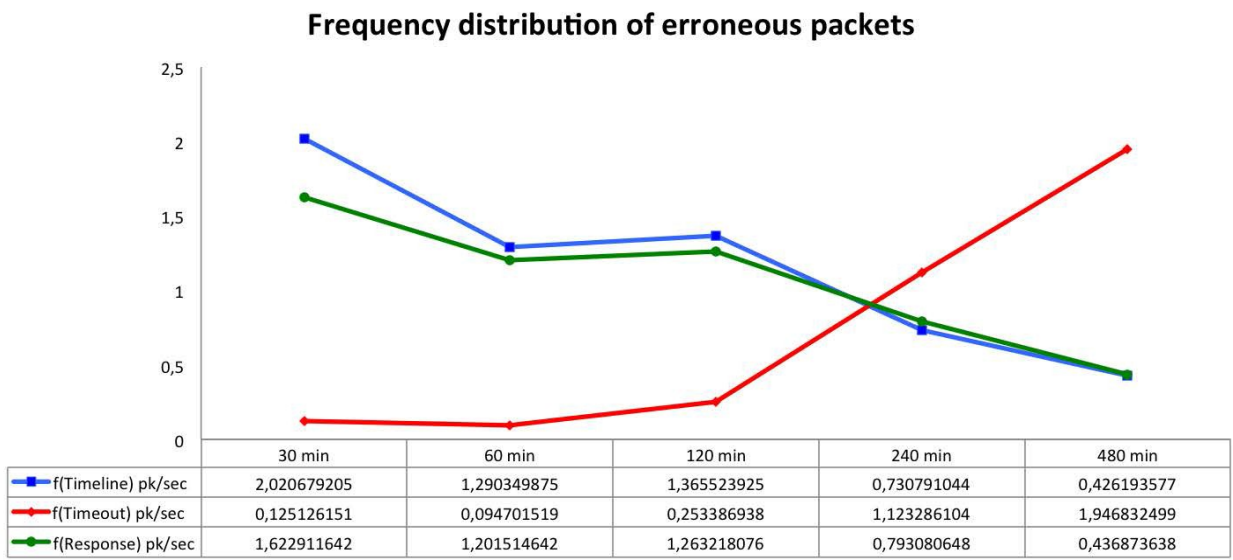

Fig. 6. Simulation results: frequency distribution of erroneous packets for Timeout, Timeline and Response

*C. Discussion*

Timeline integrity is a general concept related to the integrity security principle, that can be applied to a large number of heterogeneous cases and environment. In particular the Good Timeline property we have introduced in present paper is particularly useful in the cloud and in the federated cloud scenarios. In digital forensics analysis, having already or being able to reconstruct a good timeline is a key success factor that too often is not taken into consideration. This is probably due to the fact that most often the "Time" factor is considered as immutable in the digital world similarly to the real world. Establishing a good timeline is a fundamental factor especially in a virtualized environment and, on a larger scale, in a cloud.

Cloud forensics requires a deeper analysis of the "Good timeline" and this work with its experimental results are in our opinion a relevant forward step on that subject.

This paper presents some relevant outcomes: it stresses the importance of the problem of maintaining a good Cloud timeline for forensics purposes; it suggests the approach to log timeline problems in real time, implementing the Live Digital Forensics paradigm; it proposes an open source Java simulator to analyze the "good timeline" of a cloud environment; it analyzes in depth the security aspects, and the limits, of every peer in the cloud; it produces an high-level "cloud time model" to focus on the scenario where the timeline is compromised; it gives experimental evidence that helps to increase the common knowledge on the subject.

Present work could not delve much into specific attack scenarios and countermeasures due to time and space constraints. However, we believe the main points that can be addressed in further work are: the scalability of the specially-built simulator in Java can be improved in order to perform tests with a larger number of nodes; a distributed Java approach but also other programming languages are being evaluated; some more complex attack scenarios can be analyzed, described and addressed in a more specific future work; the probabilistic model simulating attacks and failures can be enriched in order to study and evaluate more complex attack scenarios; "threshold levels" can be extracted and studied in order to balance between service security and availability;

## VII. Conclusion

This paper contributes to the protection of timeline integrity, in particular regarding the malicious alteration of cloud virtual machine time. Our contributions help improving the resiliency of cloud hosts and guests in order to allow and even ease forensics for the cloud.

In particular, a generic cloud time model is introduced and possible attack scenarios, consequences and countermeasures are surveyed. As well as issues related to host and guest machine time security and a novel architecture for cloud host and guest time alteration detection and resilience has been introduced. A cloud load and topology simulator has been designed and implemented (soon to be open sourced) that allowed us to perform experimental activity over the CloRoFor solution we devised.

The proposed CloRoFor architecture detects timeline alteration and guarantees the validity of the forensic investigation process. Both host and guest timeline alteration scenarios are described. As regards the latter, cloud snapshot and rollback techniques have been leveraged to bring the VM back to the original unaltered state. Simulation results show the feasibility of the proposed system. Preliminary performance figures support the viability of our approach.

Future Work will be focused on distributed timeline integrity verification and on memory optimization of the implementation to increase test duration and collected data size.